\documentclass[journal]{IEEEtran}
\usepackage{cite}
\usepackage[pdftex]{graphicx}
\usepackage{amsmath}
\usepackage{algorithmic}
\usepackage{verbatim}
\usepackage{subfigure}
\usepackage{fixltx2e}
\usepackage{url}
\usepackage{array}
\usepackage{multirow}
\usepackage{multicol}
\usepackage{longtable}
\usepackage{rotating}
\usepackage{times}
\usepackage{bm}
\makeatletter
\let\NAT@parse\undefined
\makeatother
\usepackage{hyperref}  

\begin{document}

\title{HyperCI: A Higher Order Collective Influence Measure for Hypernetwork Dismantling}

\author{Dengcheng Yan, Zijian Wu, Yi Zhang, Shiqin Qu, Yiwen Zhang*, and Hong Zhong

\thanks{This work was supported by the National Natural Science Foundation of China (Grant No. U1936220,61872002) and the University Natural Science Research Project of Anhui Province (Grant No. KJ2019A0037).}

\thanks{Dengcheng Yan, Zijian Wu, Yi Zhang, Shiqin Qu, Yiwen Zhang and Hong Zhong are with the School of Computer Science and Technology, Anhui University, Hefei 230039, China (e-mail: yanzhou@ahu.edu.cn, wzjstuahu@gmail.com, zhangyi.ahu@gmail.com, qsq74708@foxmail.com, zhangyiwen@ahu.edu.cn, zhongh@ahu.edu.cn). Corresponding author: Yiwen Zhang.}
}

\markboth{Journal of \LaTeX\ Class Files,~Vol.~14, No.~8, August~2021}%
{Shell \MakeLowercase{\textit{et al.}}: HyperCI: A Higher Order Collective Influence Measure for Hypernetwork Dismantling}

\maketitle

\begin{abstract}

The connectivity of networked systems is often dependent on a small portion of critical nodes. Network dismantling studies the strategy to identify a subset of nodes the removal of which will maximally destroy the connectivity of a network and fragment it into disconnected components. However, conventional network dismantling approaches focus on simple network which models only pairwise interaction between nodes while groupwise interactions among arbitrary number of nodes are ubiquitous in networked systems like integrated circuits. Groupwise interactions modeled by hypernetwork introduce higher order connectivity patterns, which limits the application of conventional network dismantling methods on hypernetwork. In this brief, we propose HyperCI, a higher order collective influence measure for hypernetwork dismantling. It considers the node co-occurrence characteristics and higher order influence ability both introduced by hyperedges in hypernetwork. We evaluate the effectiveness of our proposed HyperCI on six real world hypernetworks including integrated circuits and citation networks and the results indicate our proposed HyperCI outperforms baseline network dismantling methods for both simple network and hypernetwork.
\end{abstract}



\section{Introduction}

\IEEEPARstart{N}{etworked} systems composed of interacting components are ubiquitous in engineering and social systems such as circuit system \cite{thulasiraman2019netcircuit} and critical infrastructure network \cite{gao2021vulpowerg,xia2008attackvul}. Recent studies \cite{li2021percolation,yan2022hitter,ma2022netrobust,liu2017recogkeynode,braunstein2016networkdismantling,ren2019gnd} have shown that their proper functioning are highly dependent on the topological properties of the underlying network structure among the interacting components, and usually the failure of a rather small portion of these components will fragment the systems into disconnected sub-modules, resulting in system functionality degradation. Thus, it is critical to identify a set of critical components in networked systems for both system protection and intentional attack. For example, in an integrated circuit composed of a large number of interconnected logic gates, some critical gates play an important bridge role in connecting the whole circuit and thus face the aging problem caused by the NBTI effect. Most conventional approaches to alleviating this problem and improving the reliability of integrated circuits such as IVC \cite{rahman2005efficient} and TG-based\cite{lin2012leakage} methods usually perform aging protection on all gates, which brings many additional area overhead and delay increase. Instead, performing aging protection on a small portion of critical gates will be more efficient. Then the remaining question becomes how to efficiently pick out these critical gates.

Fortunately, network dismantling \cite{braunstein2016networkdismantling,ren2019gnd} as an emerging field in network science aims to identify this optimal set of critical nodes in a network the removal of which will maximize the degradation of network connectivity. Then the identified critical nodes can be utilized to perform protection or attack on the networked system. As the network dismantling problem is NP-hard \cite{zhao2020dismantling} and it is computationally intractable to obtain the exact solution for large networks, many approximate approaches have been proposed from the perspectives of centrality and heuristic. Centrality-based approaches \cite{lu2016vital,morone2015ci} determine the optimal set of nodes based on nodes' centrality measure while heuristic approaches \cite{braunstein2016networkdismantling,ren2019gnd} usually design novel decycling and tree-breaking strategies for network dismantling.

\begin{figure*}
	\centering
	\subfigtopskip=0pt
	\subfigure[{The b01 Circuit}]{
		\includegraphics[scale=0.25]{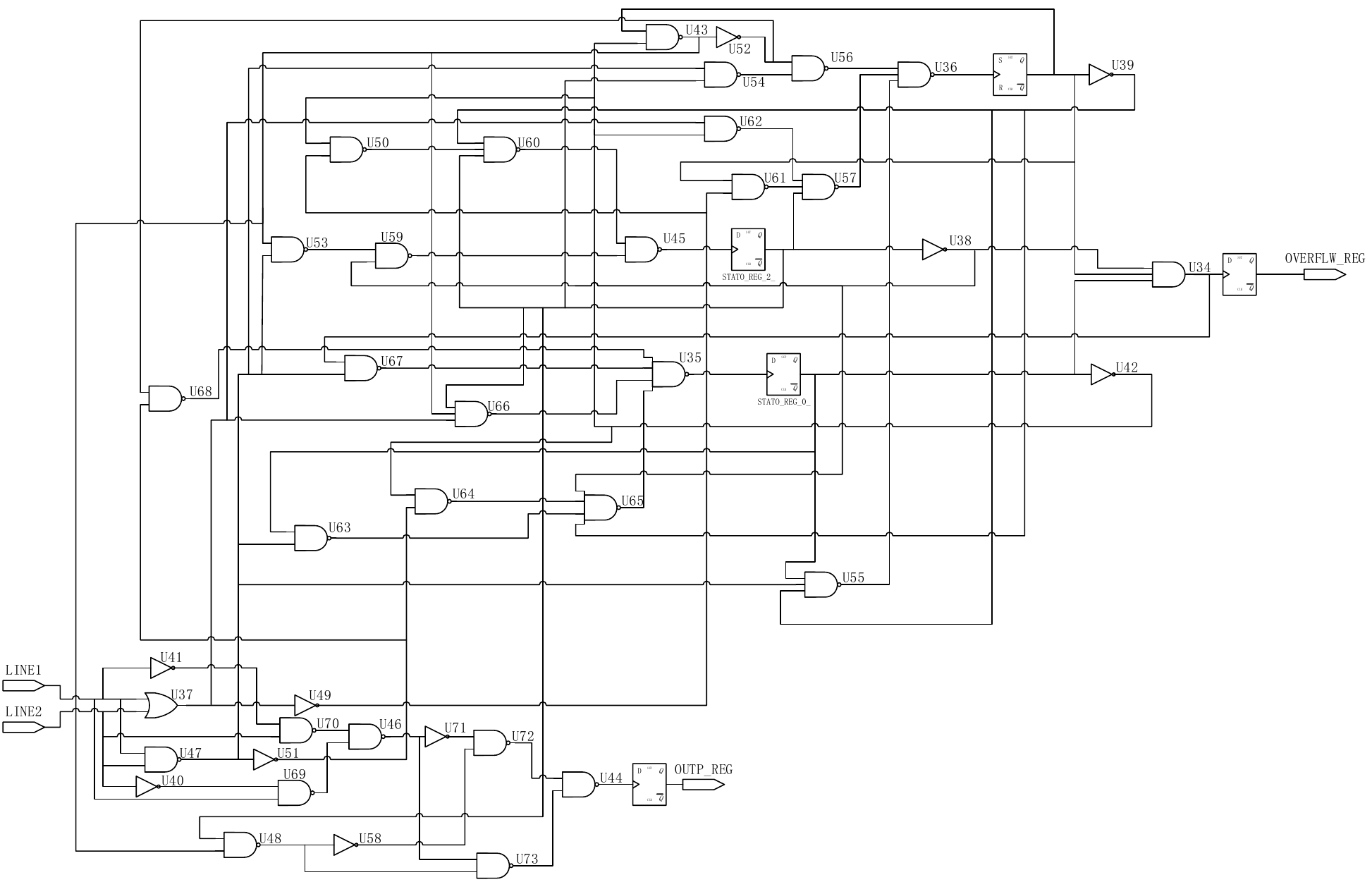}
		\label{fig:circuit}
	}
	\subfigure[Hypernetwork Representation of the b01 Circuit]{
		\includegraphics[scale=0.45]{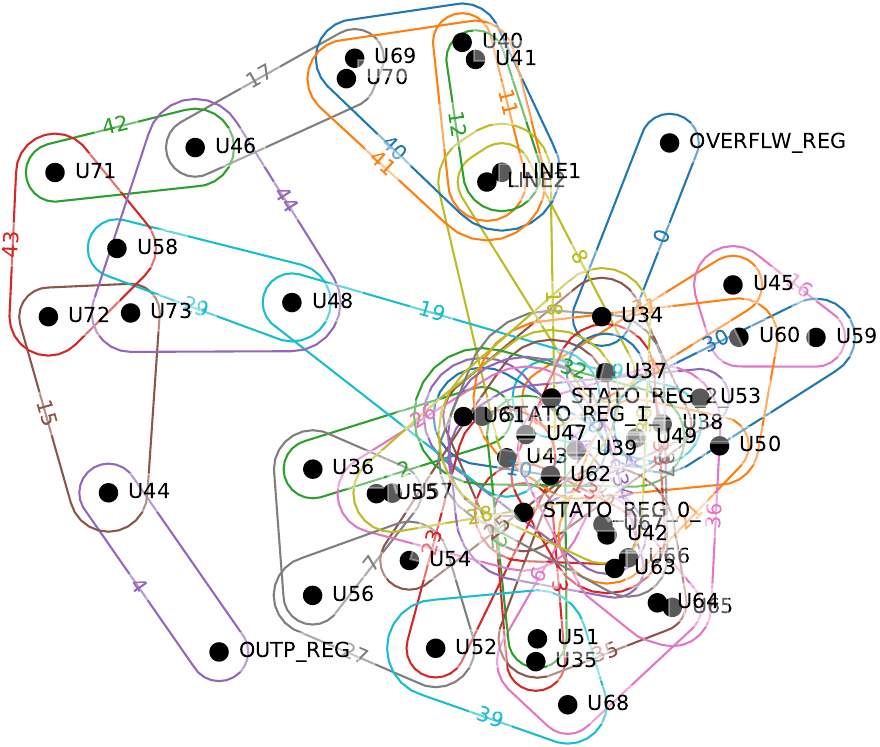}
		\label{fig:hypernetwork-rep}
	}
	\caption{The b01 circuit from ITC'99 Benchmark \cite{b01circuit} and its hypernetwork representation. Signals and logical gates in the circuit are modeled as nodes and hyperedges in its hypernetwork representation, respectively.}
	\label{fig:circuit-n-hypernetwork-rep}
\end{figure*}

However, the majority of existing network dismantling methods are limited to networks modeling pairwise interactions between two nodes while group-wise interactions among more than two nodes are also ubiquitous in many real-world systems \cite{de2020hypercontagion,zhang2016hyperd2d} and can be modeled as hypernetwork \cite{battiston2020honetworks}. A typical real-world example is the integrated circuit in which usually more than two signals are involved within one logical gate. In such a case, an integrated circuit can be modeled as a hypernetwork with signals as nodes and logical gates as hyperedges (See Fig. \ref{fig:circuit-n-hypernetwork-rep}). Group-wise interactions modeled by hyperedges introduce higher order connectivity patterns, thus the optimal dismantling of hypernetworks emerges as a distinguishing and important research topic \cite{coutinho2020hypermvc}.

A naive approach for hypernetwork dismantling is to project hypernetwork into its 2-section network form \cite{bretto2013hypergraph} by simplifying each group-wise interaction as pairwise interactions between each pair of nodes in the group, and then apply existing network dismantling methods to the 2-section network. But this approximation will obscure some structures like triangles \cite{lambiotte2019networks} which are critical for dismantling. The alternative way is to extend network dismantling approaches to hypernetwork. Intuitively, degree defined in networks as the number of edges connected to a target node can be extended as hyperdegree in hypernetworks as the number of hyperedges connected to a target node. Then hyperdegree can be utilized as a hypernetwork-specific centrality measure for hypernetwork dismantling. Tudisco et al. \cite{tudisco2021node} generalize existing network science concepts and propose node and edge centralities on hypergraphs which can be utilized for hypernetwork dismantling.

In this brief, we extend the famous CI (Collective Influence) \cite{morone2015ci} on hypernetworks and propose a novel hypernetwork-specific centrality HyperCI for hypernetwork dismantling. The HyperCI centrality takes into account not only the collective influence of nodes but also that of hyperedges which represent the higher order characteristics. Extensive experiments have been conducted on six real world datasets including both circuit and citation hypernetworks, and the results demonstrate the effectiveness of our proposed HyperCI centrality in hypernetwork dismantling.

\section{Preliminaries and Problem Definition}
In this section, we first introduce some preliminary concepts in hypernetwork and then give a formal definition of hypernetwork dismantling problem.

A hypernetwork $H$ is a collection of $N$ nodes $V=\{v_1, v_2, \ldots, v_N\}$ and $M$ hyperedges $E=\{e_1,e_2, \ldots, e_M\}$, defined as $H=(V,E)$. And each hyperedge $e_{i}=\{v_{i_1},\cdots, v_{i_k}\}$ is a subset of nodes, representing the groupwise interaction among these nodes. A hypernetwork can be mathematically represented as an incidence matrix $\bm{I} \in \{0, 1\}^{|V| \times |E|}$, and $\bm{I}_{ve}=1$ indicates node $v$ is a member of hyperedge $e$, otherwise $\bm{I}_{ve}=0$.

As in simple network, degree can be defined to measure the heterogeneous connecting ability of each node in hypernetwork. But in simple network, as each edge connects only two nodes, degree in simple network simultaneously measures the number of nodes and edges a node connects to, while this is not applicable in hypernetwork due to hyperedge's ability to connect arbitrary number of nodes. Thus, two discriminating degree measures i.e., degree $d$ and hyperdegree $d^h$ are needed to measure the number of nodes and edges a node connects to, respectively.

\begin{equation}
  \label{eq:hypernetwork_hyperdegree}
  d^h(v_i)=\sum_{j=1}^{|E|} I_{v_{i}e_{j}}
\end{equation}

\begin{equation}
  \label{eq:hypernetwork_degree}
  d(v_i)= \sum_{j=1}^{|V|} (\bm{II}^{T})_{ij} - d^h(v_i)
\end{equation}

Nodes in hypernework are connected by hyperedges and the functionality of networked systems often resides in the connectivity among nodes. Formally, the connectivity $\sigma (H)$ of a hypernetwork $H$ is defined as the ratio of the number of nodes in the giant connected component(GCC) to the total number of nodes in the whole hypernetwork,
\begin{equation}
  \label{eq:hypernetwork_connectivity}
  \sigma (H)=\frac{\left|V_{GCC}\right|}{\left|V_{H}\right|} 
\end{equation}
where $GCC$ is the giant connected component of $H$ with the most hyperedges, and $|V_{GCC}|$ and $|V_H|$ denote the number of nodes in the giant connected component and the whole hypernetwork, respectively.

Hypernetwork dismantling aims at finding an optimal set of nodes $\kappa=\left\{v_{1}, v_{2}, \cdots, v_{K}\right\}$ the removal of which will minimize the connectivity,
\begin{equation}
  \label{eq:hypernetwork_dismantling_definition}
  \mathop{min}_{\{v_{1}, v_{2}, \cdots, v_{K}\}} \sigma\left(H \backslash\left\{v_{1}, v_{2}, \cdots, v_{K}\right\}\right)
\end{equation}
where $H \backslash\left\{v_{1}, v_{2}, \cdots, v_{k}\right\}$ denotes the hypernetwork after removing nodes set $\left\{v_{1}, v_{2}, \cdots, v_{K}\right\}$ from $H$.

\section{Proposed method}
As CI in simple network measures the collective influence of a target node $v_i$ by considering the influence ability (i.e., degree $k$) of both itself and its $L$-hop neighbors $\partial Ball(v_i, L)$ in Equation (\ref{eq:CI}), a naive extension of CI to hypernetwork is to replace the degree $k$ in simple network with hyperdegree $d^h$ in hypernetwork as shown in Equation (\ref{eq:HyperCI_naive}) in which $(d^h(v_i)-1)$  and $\sum_{j \in \partial Ball(v_i, L)} (d^h(v_j)-1)$ indicate the influence ability of node $v_i$ itself and its neighbors, respectively.

\begin{equation}
  \label{eq:CI}
  CI_{L}(v_i)=(k_i-1)\sum_{j \in \partial Ball(v_i, L)} (k_j-1)
\end{equation}

\begin{equation}
  \label{eq:HyperCI_naive}
  HyperCI_{naive,L}(v_i)=(d^h(v_i)-1)\sum_{v_j \in \partial Ball(v_i, L)} (d^h(v_j)-1)
\end{equation}

\begin{figure}
	\centering
	\includegraphics[scale=0.35]{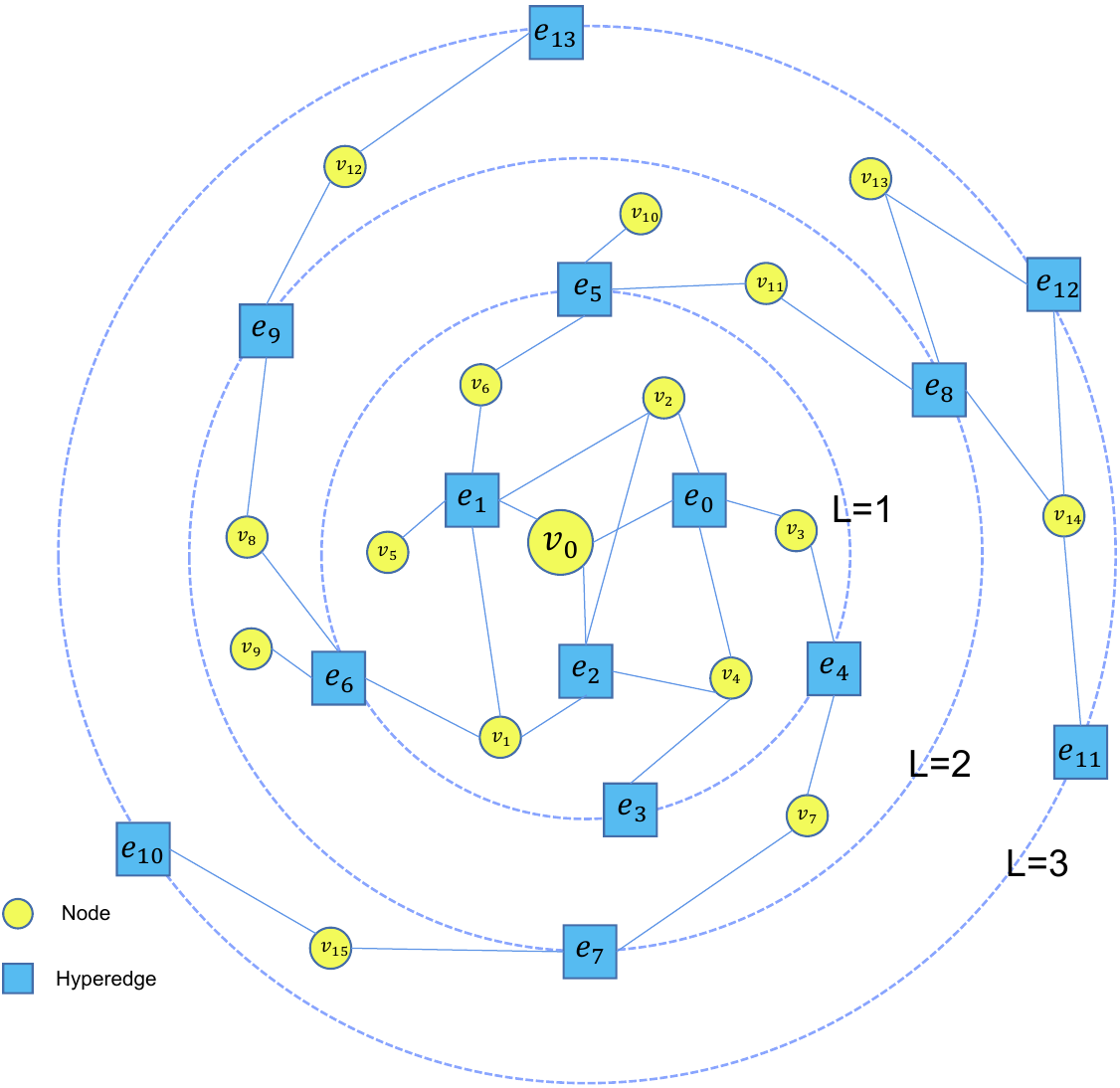}
	\caption{Illustration of the model of HyperCI algorithm. To clearly explain the influence ability of hyperedge, we utilize a bipartite representation of hypernetwork.}
	\label{fig:model-illustration}
\end{figure}

However, this naive extension of CI, on the one hand, ignores the fact that a hyperedge can connect more than two nodes and some nodes often co-occur in several common hyperedges. On the other hand, it doesn't consider the influence ability of hyperedges. We further improve HyperCI$_{naive}$ to HyperCI by considering the above two facts and a detailed illustration is shown in Fig. \ref{fig:model-illustration}.

First, we consider the case of nodes' co-occurrence in common hyperedges. As nodes co-occurring in common hyperedges enables multi paths for influence propagation and these nodes share a same influence ability which leads to the reduce of the influence ability of each individual node. Thus, we introduce a co-occurrence degree to regularize the influence ability of the target node itself as Equation (\ref{eq:node-inf}),
\begin{equation}
  inf(v_i) = \frac{d^h(v_i) - 1}{|\bigcap_{e \in \Gamma(v_i)} e|}
  \label{eq:node-inf}
\end{equation}
where $|\bigcap_{e \in \Gamma(v_i)} e|$ is the co-occurrence degree, and $\Gamma(v_i)$ represents the set of direct hyperedges of node $v_i$.
As in Fig. \ref{fig:model-illustration}, $d^h(v_0)=3$, $\Gamma(v_0)=\{e_0, e_1,e_2\}$, $e_0=\{v_0,v_2,v_3,v_4\}$, $e_1=\{v_0,v_1,v_2,v_5,v_6\}$, $e_2=\{v_0,v_1,v_2,v_4\}$, $|\bigcap_{e \in \Gamma(v_0)} e|=2$ and $inf(v_0)=1$.

Then we consider the influence ability of hyperedges. As CI and HyperCI$_{naive}$ indicate, the influence ability of $L$-hop neighbors is measured by the total number of edges/hyperedges they can connect out. In simple network, each edge can only connect out to one node while in hypernetwork each hyperedge can connect out to one or more nodes which can be consisted in different hyperedges. Thus, in hypernetwork we measure the influence ability of the neighbors of the target node by a combination of both the total number of hyperedges and the expansion capability of these hyperedges as shown in Equation (\ref{eq:neighbor-inf}),
\begin{equation}
  \label{eq:neighbor-inf}
  inf_{\Gamma}(v_i, L) = |\Gamma (v_i, L)| +  \sum_{e \in \Gamma(v_i, L)} \frac{\sum_{v \in e} |\Gamma (v) \cap \Gamma(v_i, L+1)|}{|\Psi(e, v_i)|}
\end{equation}
where $\Gamma (v_i, L)$ is the set of $v_i$'s $L$-hop neighboring hyperedges and $\Psi(e, v_i) = \{v | v \in e \cap e^{\prime}, e^{\prime} \in \Gamma (v_i, L+1)\}$ is the number of nodes hyperedge $e$ can connect out.
As shown in Fig. \ref{fig:model-illustration}, $\Gamma (v_0, 2) = \{e_7,e_8,e_9\}$, $\Gamma (v_0, 3) = \{e_{10},e_{11},e_{12},e_{13}\}$, $\sum_{v \in e_{8}} |\Gamma (v) \cap \Gamma(v_0, 3)| = |\Gamma (v_{11}) \cap \Gamma(v_0, 3)| + |\Gamma (v_{13}) \cap \Gamma(v_0, 3)|+ |\Gamma (v_{14}) \cap \Gamma(v_0, 3)| = | \{e_5, e_8\} \cap \{e_{10},e_{11},e_{12},e_{13}\}| + | \{e_8, e_{12}\} \cap \{e_{10},e_{11},e_{12},e_{13}\}|+ | \{e_8, e_{11},e_{12}\} \cap \{e_{10},e_{11},e_{12},e_{13}\}| = 3$, $|\Psi(e_{8}, v_0)| = |\{v | v \in e_{8} \cap e^{\prime}, e^{\prime} \in \Gamma (v_0, 3)\}| = |\{v_{13},v_{14}\}| = 2$, thus $\sum_{v \in e_{8}} |\Gamma (v) \cap \Gamma(v_0, 3)| = 1.5$, then we have $\sum_{v \in e_{7}} |\Gamma (v) \cap \Gamma(v_0, 3)| = 1$, $\sum_{v \in e_{9}} |\Gamma (v) \cap \Gamma(v_0, 3)| = 1$, so, we can calculate $inf_{\Gamma}(v_0, 2)=|\Gamma (v_0, 2)|+1.5+1+1=6.5$

Besides the influence ability of the $L$-hop hyperedges, the number of nodes contained in these neighboring hyperedges also indicates the influence of the target node $v_i$. Thus, we introduce the neighboring node influence ability as Equation (\ref{eq:neighboring-node-inf}), 
\begin{equation}
  \label{eq:neighboring-node-inf}
  inf_N(v_i, L) = |\{v | v \in \bigcup_{e \in \Gamma (v_i, L)} e\}|
\end{equation}

Finally, we get the HyperCI of a target node $v_i$ by multiplying these different influence ability as shown in Equation (\ref{eq:HyperCI}).
\begin{equation}
  \label{eq:HyperCI}
  HyperCI(v_i, L) = inf(v_i) \times inf_{\Gamma}(v_i, L) \times inf_N(v_i, L)
\end{equation}

In Fig. \ref{fig:model-illustration}, $inf_N(v_0, 2)=|\{v_7,v_{15},v_8,v_{12},v_{11},v_{13},v_{14}\}|=7$, last, we can get $HyperCI(v_0, 2)=1*6.5*7=45.5$

\section{Experiment}

In this section, we will conduct experiments on six real-word hypernetwork to demonstrate the performance of our proposed $HyperCI$ on the task of hypernetwork dismantling.

\subsection{Experimental Settings}

We evaluate the performance of our proposed metric on six real-world hypernetworks from different domains including citation hypernetwork \cite{ley2002dblp, benson2018simplicial, sinha2015overview} such as Cora, MAG and DBLP from scientific collaboration and logic circuit hypernetworks such c1908, c3540, s5378 from VLSI design \cite{brglez1985neural,brglez1989combinational}. For citation hypernetworks authors are modeled as nodes and co-authorships are modeled as hyperedges, and for logic circuit hypernetworks signals are modeled as nodes and each logic gate containing multiple signals is modeled as a hyperedges. The statistical information of these hypernetworks is listed in Table \ref{tab:dataset}.

\begin{table*}[!htbp]  
    \centering
    \caption{Statistical information of the datasets.}
    \begin{tabular}{|c|c|c|c|c|c|c|}  
    \hline
    Dataset &   \#Nodes & \#Hyperedges &  Avg. Node Degree &  Avg. Hyperedge Size & Avg. Shortest Path Length (Node) & Avg. Shortest Path Length (Edge) \\  
    \hline
    Cora     &    1676 &    463  &    1.66 &    6.00&   5.27&   4.98\\ \hline
    DBLP     &    4695 &    2561 &    3.06 &    5.61&   4.59&   3.76\\ \hline
    MAG      &    4039 &    1984 &    1.70 &    3.47&   14.75&  14.44\\ \hline
    c1908    &    913  &    913  &    2.64 &    2.64&   8.61&   8.52 \\ \hline
    c3540    &    1719 &    1719 &    2.71 &    2.71&   7.43&   7.14 \\ \hline
    s5378    &    2993 &    2993 &    2.46 &    2.46&   9.96&   9.79 \\ \hline
    \end{tabular}
    \label{tab:dataset}
\end{table*}

To demonstrate the effectiveness of our proposed method HyperCI on hypernetwork dismantling, we compare it with dismantling methods designed for both simple network and hypernetwork, including Highest Degree (HD), Adaptive Highest Degree (HDA), CI\cite{morone2015ci} and GND\cite{ren2019gnd} for simple network dismantling, and Highest Hyperdegree (HHD), Adaptive Highest Hyperdegree (HHDA), COMMP\cite{tudisco2021node}, subTSSH\cite{antelmi2021social} and HyperCI$_{naive}$ for hypernetwork dismantling. It is worth noting that dismantling methods designed for simple network are often applied to hypernetwork dismantling by transforming hypernetwork into its simple network form, i.e., 2-section network for simplicity. In this experiment, we also utilize this kind of transformation. The dismantling process is conducted with the following steps: (1) A score is calculated for each node by a given dismantling method; (2) The top $1\%$ nodes with the highest scores are removed from the hypernetwork; (3) Reconstruct the remaining hypernetwork by deleting hyperedges containing no nodes and merging hyperedges containing same node sets; (4) Repeat the above steps until all the remaining nodes are disconnected. This dismantling process is applied for all the dismantling methods in this experiment except HD and HHD as they are not adaptive. For HD and HHD, the score is only calculated once at the beginning of the dismantling process and nodes are removed according to the score in descending order.

\subsection{Evaluation Metric}

The performance of HyperCI and the baselines is evaluated against the accumulated normalized connectivity(ANC)\cite{schneider2011anc} metric as shown in Equation (\ref{eq:ANC}),
\begin{equation}
  ANC(\kappa)=\frac{1}{K} \sum_{k=1}^{K} \frac{\sigma\left(H \backslash\left\{v_{1}, v_{2}, \cdots, v_{k}\right\}\right)}{\sigma(H)}
  \label{eq:ANC}
\end{equation}
where $K$ is the size of the removed node set $\kappa=\left\{v_{1}, v_{2}, \cdots, v_{K}\right\}$ and $\sigma ()$ is a connectivity measure as defined in Equation (\ref{eq:hypernetwork_connectivity}).

\subsection{Results}

The overall performance of our proposed HyperCI and the baselines is shown in Table \ref{tab:exp-performance}. From the results, we can find that the simple higher order extension of CI, i.e., HyperCI$_{naive}$ can obtain a significant improvement in hypernetwork dismantling, and HyperCI performs even better. This indicates the effectiveness of taking into account the characteristics of higher order interactions in hypernetwork including both node co-occurrence in hyperedges and influence ability of hyperedges. Moreover, comparing dismantling methods designed for hypernetwork with those for simple network, we can also find the former ones often perform better than the latter ones, which strengthens the motivation to utilize the characteristics of higher order interactions for hypernetwork dismantling. Comparing HyperCI itself with different parameter $L$, it can be found that for citation hypernetworks (Cora, DBLP, MAG) HyperCI with smaller $L$ performs better while for logic circuit hypernetworks (c1908, c3540, s5378) HyperCI with larger $L$ performs better. This is because the average length of shortest path of the logic circuit hypernetworks are larger than the citation hypernetworks (except MAG) and farther neighbors are needed to measure the influence of target node. As for MAG, though its average length of shortest path is even larger, its average node degree is small, making it an extremely sparse hypernetwork, thus smaller $L$ is better.

\begin{table*}
    \centering
    \caption{The overall performance.}
    \label{tab:exp-performance}
    \begin{tabular}{|c|c|c|c|c|c|c|c|c|c|c|c|c|}  
    \hline
    \multirow{2}{*}{Datasets} &   \multirow{2}{*}{HD}   &    \multirow{2}{*}{HDA}   &   \multirow{2}{*}{CI}      &   \multirow{2}{*}{GND}    &    \multirow{2}{*}{HHD}    &    \multirow{2}{*}{HHDA}       & 
    \multirow{2}{*}{COMMP}  & \multirow{2}{*}{subTSSH}    & \multicolumn{2}{c|}{HyperCI$_{naive}$} & \multicolumn{2}{c|}{HyperCI}  \cr\cline{10-13} 
      &      &         &         &        &          &           & 
      &      & L=1 & L=2 & L=1 & L=2  \\
    \hline
    Cora  & 0.1548 &  0.1370 &  0.1153 &  0.1344 &  0.0996 &  0.0976 &  0.0985 &  0.1256 & 0.0923 &  0.0756 &  \textbf{0.0561} &  \underline{0.0614} \\ \hline

    DBLP  & 0.1455 &  0.1380 &  0.1358 &  0.2705 &  0.1388 &  0.1236 &  0.1389 &  0.1491 & 0.1235 &  0.1135 &  \textbf{0.1042} &  \underline{0.1056} \\ \hline

    MAG   & 0.0566 &  0.0376 &  0.0385 &  0.0775 &  0.0367 &  0.0373 &  0.0351 &  0.0434 & 0.0258 &  0.0231 &  \textbf{0.0169} &  \underline{0.0205} \\ \hline
    
    c1908 & 0.2064 &  0.1547 &  0.1406 &  0.1143 &  0.1564 &  0.1204 &  0.1042 &  0.1518 & 0.0991 &  0.0994 &  \underline{0.0930} &  \textbf{0.0795} \\ \hline
    
    c3540 & 0.2194 &  0.1676 &  0.1584 &  0.3881 &  0.1582 &  0.1484 &  0.1383 &  0.1822 & 0.1282 &  0.1262 &  \underline{0.1168} &  \textbf{0.1167} \\ \hline
    
    s5378 & 0.1902 &  0.1568 &  0.1638 &  0.1845 &  0.1560 &  0.1353 &  0.1429 &  0.1508 & 0.1308 &  \underline{0.1167} &  0.1302 &  \textbf{0.1064} \\ \hline
    \end{tabular}
\end{table*}

\begin{figure*}
	\centering
	\subfigtopskip=0pt
	\subfigure[Cora]{
		\includegraphics[scale=0.35]{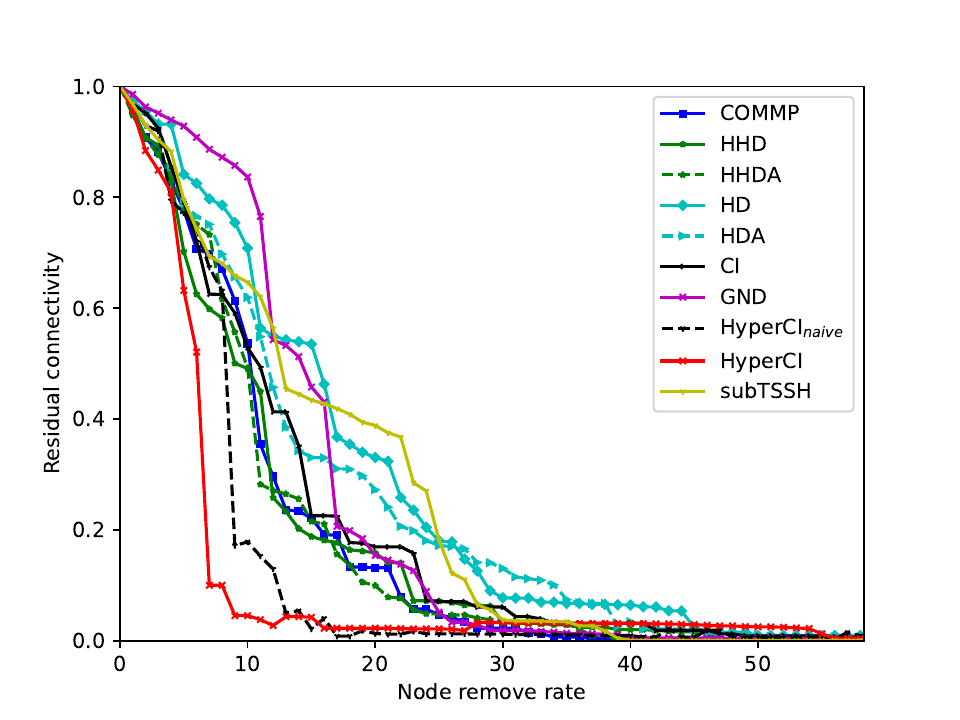}
		\label{4a}
	}
	\subfigure[DBLP]{
		\includegraphics[scale=0.35]{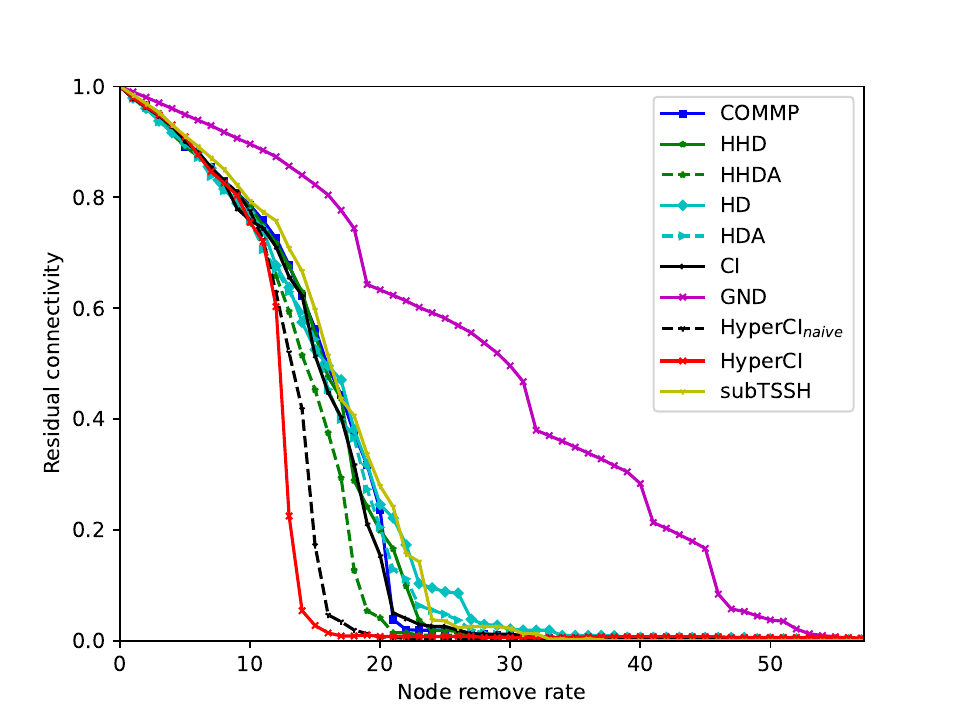}
		\label{4b}
	}
	\subfigure[MAG]{
		\includegraphics[scale=0.35]{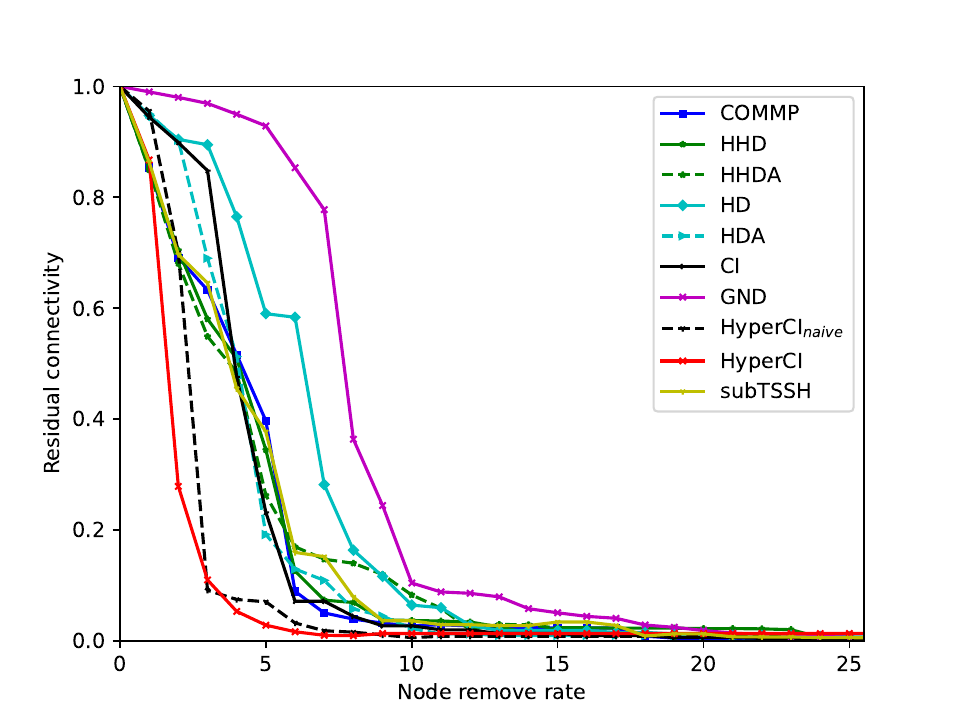}
		\label{4e}
	}
	\subfigure[c1908]{
		\includegraphics[scale=0.35]{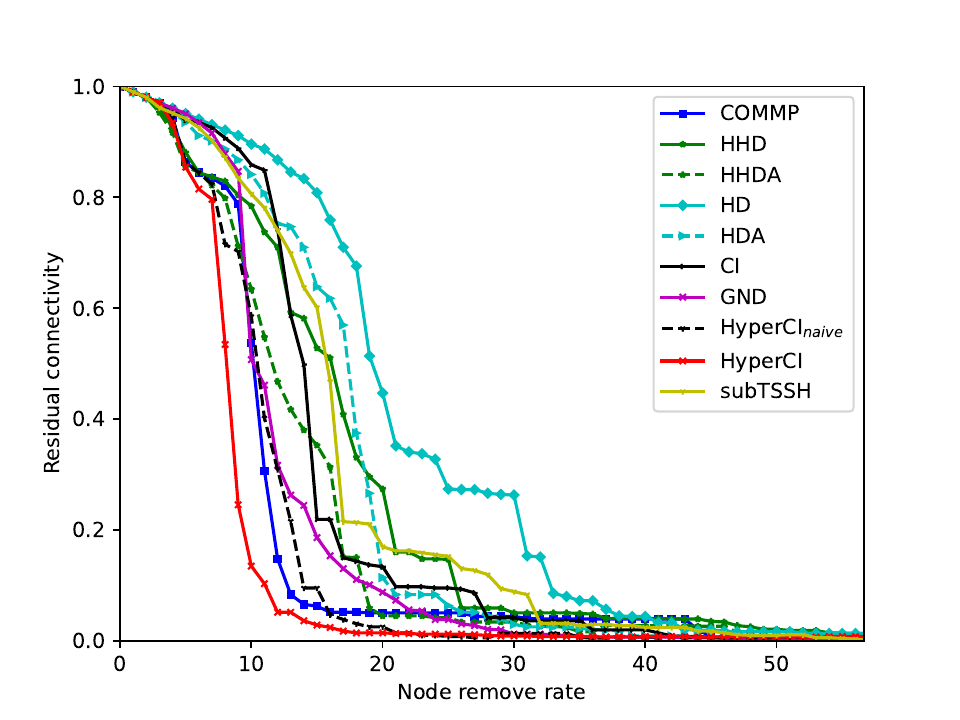}
		\label{4d}
	}
	\subfigure[c3540]{
		\includegraphics[scale=0.35]{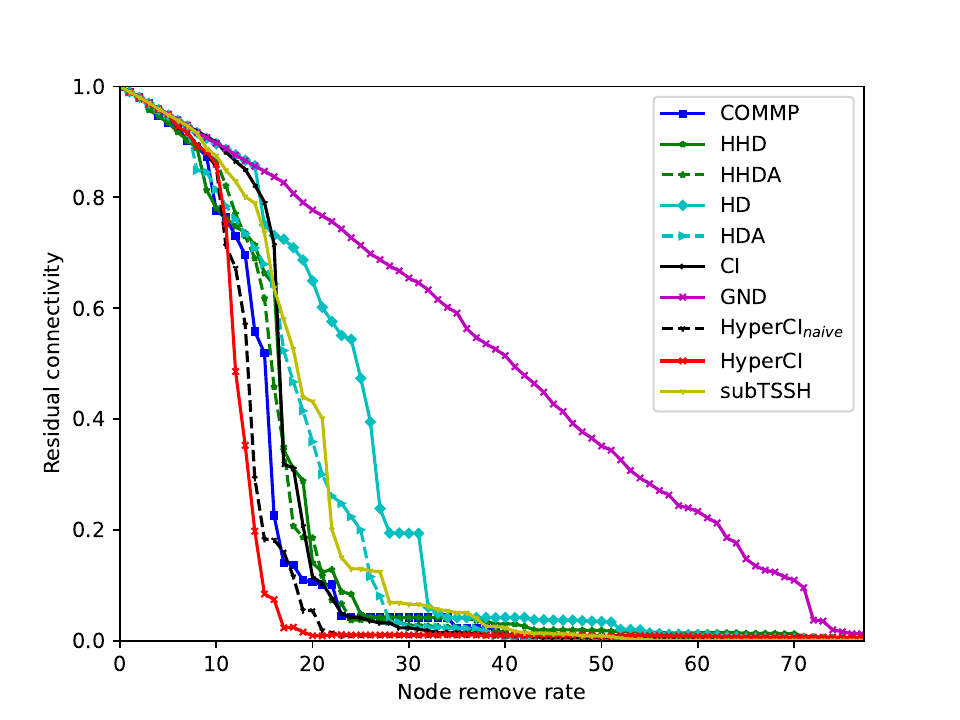}
		\label{4e}
	}
	\subfigure[s5378]{
		\includegraphics[scale=0.35]{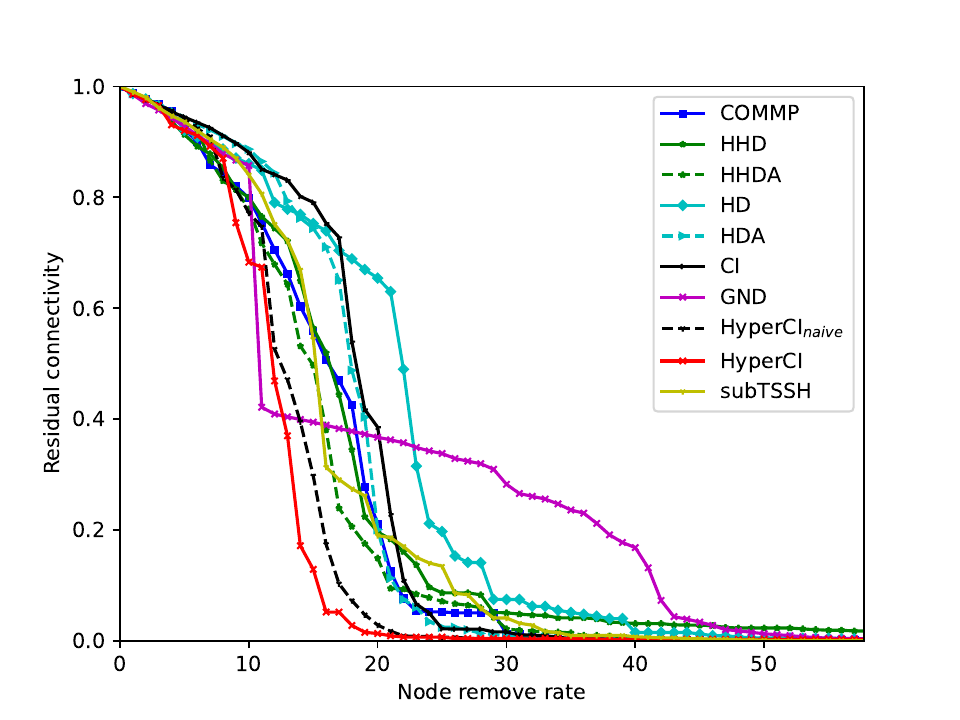}
		\label{4e}
	}
	\caption{The detailed ANC curve.}
	\label{fig:ANC-curve}
\end{figure*}

To show the dismantling process more detailedly, we plot the residual connectivity after each dismantling step as shown in Fig. \ref{fig:ANC-curve}. It can be seen from the figure that the removal of only a small portion of nodes will significantly downgrade the connectivity. This will provide us with a strong motivation to perform  protection on only a small portion of gates and signals to alleviate aging  problem in integrated circuit instead of all the gates and signals.

\section{Conclusions}

In this brief, we study the hypernetwork dismantling problem and propose HyperCI by extending CI with the consideration of higher order interactions among nodes. Experiments on real-world hypernetworks, including scientific collaboration and integrated circuit demonstrate the effectiveness of our proposed HyperCI. Our proposed method provides an efficient inspiration to perform aging protection on integrated circuits by strengthening only a small portion of critical circuit components.

\bibliographystyle{IEEEtran}
\bibliography{HyperCI}

\begin{thebibliography}{10}
\providecommand{\url}[1]{#1}
\csname url@samestyle\endcsname
\providecommand{\newblock}{\relax}
\providecommand{\bibinfo}[2]{#2}
\providecommand{\BIBentrySTDinterwordspacing}{\spaceskip=0pt\relax}
\providecommand{\BIBentryALTinterwordstretchfactor}{4}
\providecommand{\BIBentryALTinterwordspacing}{\spaceskip=\fontdimen2\font plus
\BIBentryALTinterwordstretchfactor\fontdimen3\font minus
  \fontdimen4\font\relax}
\providecommand{\BIBforeignlanguage}[2]{{%
\expandafter\ifx\csname l@#1\endcsname\relax
\typeout{** WARNING: IEEEtran.bst: No hyphenation pattern has been}%
\typeout{** loaded for the language `#1'. Using the pattern for}%
\typeout{** the default language instead.}%
\else
\language=\csname l@#1\endcsname
\fi
#2}}
\providecommand{\BIBdecl}{\relax}
\BIBdecl

\bibitem{thulasiraman2019netcircuit}
K.~Thulasiraman, M.~Yadav, and K.~Naik, ``Network science meets circuit theory:
  Resistance distance, kirchhoff index, and foster’s theorems with
  generalizations and unification,'' \emph{IEEE Trans. Circuits Syst. I-Regul.
  Pap.}, vol.~66, no.~3, pp. 1090--1103, 2019.

\bibitem{gao2021vulpowerg}
X.~Gao, C.~Pu, and L.~Li, ``Vulnerability assessment of power grids against
  cost-constrained hybrid attacks,'' \emph{IEEE Trans. Circuits Syst.
  II-Express Briefs}, vol.~68, no.~4, pp. 1477--1481, 2021.

\bibitem{xia2008attackvul}
Y.~Xia and D.~J. Hill, ``Attack vulnerability of complex communication
  networks,'' \emph{IEEE Trans. Circuits Syst. II-Express Briefs}, vol.~55,
  no.~1, pp. 65--69, 2008.

\bibitem{li2021percolation}
M.~Li, R.-R. Liu, L.~L{\"u}, M.-B. Hu, S.~Xu, and Y.-C. Zhang, ``Percolation on
  complex networks: Theory and application,'' \emph{Phys. Rep.}, vol. 907, pp.
  1--68, 2021.

\bibitem{yan2022hitter}
D.~Yan, W.~Xie, Y.~Zhang, Q.~He, and Y.~Yang, ``Hypernetwork dismantling via
  deep reinforcement learning,'' \emph{IEEE Trans. Netw. Sci. Eng.}, 2022.

\bibitem{ma2022netrobust}
W.~Ma, J.~Fang, and J.~Wu, ``Analyzing robustness of complex networks against
  incomplete information,'' \emph{IEEE Trans. Circuits Syst. II-Express
  Briefs}, vol.~69, no.~5, pp. 2523--2527, 2022.

\bibitem{liu2017recogkeynode}
B.~Liu, Z.~Li, X.~Chen, Y.~Huang, and X.~Liu, ``Recognition and vulnerability
  analysis of key nodes in power grid based on complex network centrality,''
  \emph{IEEE Trans. Circuits Syst. II-Express Briefs}, vol.~65, no.~3, pp.
  346--350, 2017.

\bibitem{braunstein2016networkdismantling}
A.~Braunstein, L.~Dall’Asta, G.~Semerjian, and L.~Zdeborov{\'a}, ``Network
  dismantling,'' \emph{Proc. Natl. Acad. Sci.}, vol. 113, no.~44, pp.
  12\,368--12\,373, 2016.

\bibitem{ren2019gnd}
X.-L. Ren, N.~Gleinig, D.~Helbing, and N.~Antulov-Fantulin, ``Generalized
  network dismantling,'' \emph{Proc. Natl. Acad. Sci.}, vol. 116, no.~14, pp.
  6554--6559, 2019.

\bibitem{rahman2005efficient}
H.~Rahman and C.~Chakrabarti, ``An efficient control point insertion technique
  for leakage reduction of scaled cmos circuits,'' \emph{IEEE Trans. Circuits
  Syst. II-Express Briefs}, vol.~52, no.~8, pp. 496--500, 2005.

\bibitem{lin2012leakage}
C.~Lin, C.-H. Lin, and K.-H. Li, ``Leakage and aging optimization using
  transmission gate-based technique,'' \emph{IEEE Trans. Comput-Aided Des.
  Integr. Circuits Syst.}, vol.~32, no.~1, pp. 87--99, 2013.

\bibitem{zhao2020dismantling}
D.~Zhao, S.~Yang, X.~Han, S.~Zhang, and Z.~Wang, ``Dismantling and vertex cover
  of network through message passing,'' \emph{IEEE Trans. Circuits Syst.
  II-Express Briefs}, vol.~67, no.~11, pp. 2732--2736, 2020.

\bibitem{lu2016vital}
L.~L{\"u}, D.~Chen, X.-L. Ren, Q.-M. Zhang, Y.-C. Zhang, and T.~Zhou, ``Vital
  nodes identification in complex networks,'' \emph{Phys. Rep.}, vol. 650, pp.
  1--63, 2016.

\bibitem{morone2015ci}
F.~Morone and H.~A. Makse, ``Influence maximization in complex networks through
  optimal percolation,'' \emph{Nature}, vol. 524, no. 7563, pp. 65--68, 2015.

\bibitem{de2020hypercontagion}
G.~F. de~Arruda, G.~Petri, and Y.~Moreno, ``Social contagion models on
  hypergraphs,'' \emph{Phys. Rev. Res.}, vol.~2, no.~2, p. 023032, 2020.

\bibitem{zhang2016hyperd2d}
H.~Zhang, L.~Song, and Z.~Han, ``Radio resource allocation for device-to-device
  underlay communication using hypergraph theory,'' \emph{IEEE Trans. Wirel.
  Commun.}, vol.~15, no.~7, pp. 4852--4861, 2016.

\bibitem{battiston2020honetworks}
F.~Battiston, G.~Cencetti, I.~Iacopini, V.~Latora, M.~Lucas, A.~Patania, J.-G.
  Young, and G.~Petri, ``Networks beyond pairwise interactions: structure and
  dynamics,'' \emph{Phys. Rep.}, vol. 874, pp. 1--92, 2020.

\bibitem{coutinho2020hypermvc}
B.~C. Coutinho, A.-K. Wu, H.-J. Zhou, and Y.-Y. Liu, ``Covering problems and
  core percolations on hypergraphs,'' \emph{Phys. Rev. Lett.}, vol. 124,
  no.~24, p. 248301, 2020.

\bibitem{b01circuit}
S.~Davidson, ``{Notes on ITC'99 Benchmarks},''
  https://www.cerc.utexas.edu/itc99-benchmarks/bendoc1.html.

\bibitem{bretto2013hypergraph}
A.~Bretto, \emph{Hypergraph Theory: An Introduction}.\hskip 1em plus 0.5em
  minus 0.4em\relax Springer International Publishing, 2013.

\bibitem{lambiotte2019networks}
R.~Lambiotte, M.~Rosvall, and I.~Scholtes, ``From networks to optimal
  higher-order models of complex systems,'' \emph{Nat. Phys.}, vol.~15, no.~4,
  pp. 313--320, 2019.

\bibitem{tudisco2021node}
F.~Tudisco and D.~Higham, ``Node and edge eigenvector centrality for
  hypergraphs,'' \emph{Commun. Phys.}, vol.~4, no.~1, p. 201, 2021.

\bibitem{ley2002dblp}
M.~Ley, ``The dblp computer science bibliography: Evolution, research issues,
  perspectives,'' in \emph{International Symposium on String Processing and
  Information Retrieval}, 2002, pp. 1--10.

\bibitem{benson2018simplicial}
A.~R. Benson, R.~Abebe, M.~T. Schaub, A.~Jadbabaie, and J.~Kleinberg,
  ``Simplicial closure and higher-order link prediction,'' \emph{Proc. Natl.
  Acad. Sci.}, vol. 115, no.~48, pp. E11\,221--E11\,230, 2018.

\bibitem{sinha2015overview}
A.~Sinha, Z.~Shen, Y.~Song, H.~Ma, D.~Eide, B.-J. Hsu, and K.~Wang, ``An
  overview of microsoft academic service (mas) and applications,'' in
  \emph{Proceedings of the 24th International Conference on World Wide Web},
  2015, pp. 243--246.

\bibitem{brglez1985neural}
F.~Brglez, ``A neural netlist of 10 combinational benchmark circuits,''
  \emph{Proc. IEEE ISCAS: Special Session on ATPG and Fault Simulation}, pp.
  151--158, 1985.

\bibitem{brglez1989combinational}
F.~Brglez, D.~Bryan, and K.~Kozminski, ``Combinational profiles of sequential
  benchmark circuits,'' in \emph{IEEE International Symposium on Circuits and
  Systems}, 1989, pp. 1929--1934.

\bibitem{antelmi2021social}
A.~Antelmi, G.~Cordasco, C.~Spagnuolo, and P.~Szufel, ``Social influence
  maximization in hypergraphs,'' \emph{Entropy}, vol.~23, no.~7, p. 796, 2021.

\bibitem{schneider2011anc}
C.~M. Schneider, A.~A. Moreira, J.~S. Andrade, S.~Havlin, and H.~J. Herrmann,
  ``Mitigation of malicious attacks on networks,'' \emph{Proc. Natl. Acad.
  Sci.}, vol. 108, no.~10, pp. 3838--3841, 2011.

\end{thebibliography}

\end{document}